# Spin torque study of the spin Hall conductivity and spin diffusion length in platinum thin films with varying resistivity

Minh-Hai Nguyen[1], D. C. Ralph[1,2], R. A. Buhrman[1*]

[1]Cornell University, Ithaca, New York 14853, USA

[2]Kavli Institute at Cornell, Ithaca, New York 14853, USA

## ABSTRACT

We report measurements of the spin torque efficiencies in perpendicularly-magnetized Pt/Co bilayers where the Pt resistivity $\rho_{Pt}$ is strongly dependent on thickness $t_{Pt}$. The damping-like spin Hall torque efficiency per unit current density, $\xi_{DL}^j$, varies significantly with $t_{Pt}$, exhibiting a peak value $\xi_{DL}^j = 0.12$ at $t_{Pt}$ = 2.8 - 3.9 nm. In contrast, $\xi_{DL}^j / \rho_{Pt}$ increases monotonically with $t_{Pt}$ and saturates for $t_{Pt}$ > 5 nm, consistent with an intrinsic spin Hall effect mechanism, in which $\xi_{DL}^j$ is enhanced by an increase in $\rho_{Pt}$. Assuming the Elliott-Yafet spin scattering mechanism dominates we estimate that the spin diffusion length $\lambda_s = (0.77 \pm 0.08) \times 10^{-15}$ $\Omega m^2 / \rho_{Pt}$.

[*] rab8@cornell.edu

The spin Hall effect (SHE) [1–3], in which a transverse spin current density $j_{SHE}$ is induced by a longitudinal charge current density $j_e$ and whose strength is characterized by the spin Hall ratio $\theta_{SH} \equiv (2e/\hbar) j_{SHE} / j_e$, has recently drawn much attention because of its promise for spintronics applications [4–13]. Mechanisms which might give rise to the SHE [14,15] include the intrinsic SHE [1,16], side-jump scattering [17] and skew scattering [18]. Two common methods to quantify the strength of the SHE are to employ ferromagnet/normal metal (FM/NM) bilayers and either (1) detect the spin transfer torque that the SHE-induced spin current from the NM layer exerts on the magnetization of the adjacent FM layer [19,20], or (2) use spin pumping to inject a spin current from the FM to the NM and detect the electric current in the NM layer that is induced by the inverse SHE (ISHE) [21–23]. In the former case due to spin backflow (SBF) at the FM/NM interface [24,25] and/or enhanced spin scattering at the interface (spin memory loss or SML) [26], only a portion $j_s^{NM|FM}$ of the SHE-induced spin current $j_{SHE}$ is absorbed in the FM layer, and that reduces the damping-like (DL) spin Hall (SH) torque efficiency per unit current density $\xi_{DL}^j \equiv (2e/\hbar) j_s^{NM|FM} / j_e = T_{int} \theta_{SH}$ to be less than $\theta_{SH}$, where $T_{int} = j_s^{NM|FM} / j_{SHE}$ ($<1$) is the interfacial spin transparency. SBF and/or SML can similarly reduce the strength of spin-pumping/ISHE signals.

Large values of $\xi_{DL}^j$ have been reported for Pt [19,27–31], beta-Ta [19] and beta-W [4]. Special attention has been paid to Pt because its relatively low resistivity compared to the other SH materials would be beneficial for reducing Ohmic losses in applications. Values of $\xi_{DL}^j$ for Pt have been reported spanning the range 0.06 - 0.12 [19,27–29], depending on the FM/Pt interface [31], and are usually accompanied by a relatively small field-like (FL) torque efficiency whose magnitude and sign vary with the interface, FM magnetic anisotropy and temperature [29,32–36]. From an analysis of SBF based on a spin diffusion model [24,25], these $\xi_{DL}^j$ results indicate that the underlying internal value of $\theta_{SH}$ for Pt is ~ 0.2 or even larger [28,29,31]. However, the determination of $\theta_{SH}$ from $\xi_{DL}^j$ using the spin diffusion model requires an accurate value of the spin diffusion length $\lambda_s$, and in the case of Pt that value has long been controversial. Measurements by different techniques, at low and room temperatures, have reported a wide range, 1 - 11 nm, for $\lambda_s$ in Pt [21–23,37–48]. Those measurements will be reviewed along with our analysis later in this Letter.

Here we report that $\xi_{DL}^j$ has a strong, unexpected dependence on Pt thin film thickness $t_{Pt}$ in perpendicularly-magnetized Pt/Co bilayers, as measured by the harmonic response (HR) technique [20,29]. In particular we report that $\xi_{DL}^j$ exhibits a peak at $t_{Pt} = 2.8 - 3.9$ nm and gradually decreases at larger Pt thickness. This behavior is counter to the common expectation, reported in prior experiments with different layer structures [38,40,45], that $\xi_{DL}^j$ should simply increase and saturate at a maximum value as $t_{Pt}$ exceeds the spin diffusion length $\lambda_s$ in Pt. Our interpretation of our result is that the spin Hall ratio is linearly dependent on the Pt resistivity $\rho_{Pt}$, which in turn varies approximately inversely with $t_{Pt}$ in our samples in the thin Pt limit, $t_{Pt} \leq 4$ nm, due to strong diffusive scattering at the Pt interface(s). We observe that the spin-torque efficiency *per unit applied electric field* $\xi_{DL}^E = \xi_{DL}^j / \rho_{Pt}$ increases monotonically with $t_{Pt}$ and saturates at $t_{Pt} \approx 5$ nm. This is consistent with a spin Hall conductivity $\sigma_{SH}$ that is independent of $\rho_{Pt}$, which indicates that the intrinsic SHE (and/or side-jump scattering) determines the spin Hall ratio in our Pt films. The variation of $\xi_{DL}^E$ with $t_{Pt}$ is consistent with an *effective* $\lambda_s^{\text{eff}} = 2.0 \pm 0.1$ nm, but this determination neglects the fact that spin relaxation in Pt is predicted to be dominated by the Elliott-Yafet (E-Y) mechanism [49,50], so that $\lambda_s$ should scale linearly with $1/\rho_{\text{Pt}}$ and therefore the spin diffusion length should depend on $t_{\text{Pt}}$ in our samples, as well. We find that an analysis that assumes that $\lambda_s \rho_{\text{Pt}}$ is a constant in our bilayer samples fits the experimental results well, and from the fit we obtain $\lambda_s \rho_{\text{Pt}} = (0.77 \pm 0.08) \times 10^{-15}$ $\Omega \cdot \text{m}^2$. As discussed below, taking into account that $\lambda_s$ should scale $\propto 1/\rho_{\text{Pt}}$ would appear to resolve a prolonged controversy regarding the values of $\lambda_s$ obtained from various SHE and ISHE experiments.

We studied multilayer samples consisting of *substrate*/Ta(1)/Pt($t_{\text{Pt}}$)/Co(1)/MgO(2)/Ta(1) (numbers in parentheses are thicknesses in nm) grown on oxidized Si substrates by sputter-deposition in a vacuum of $<1.0 \times 10^{-7}$ Torr. The Ta(1) seeding layer resulted in a smoother multilayer [51,52] and enhanced perpendicular magnetic anisotropy (PMA) of the Co. The Pt thickness $t_{\text{Pt}}$, as averaged over the sample area, was varied in fine steps from 1.2 nm to 15 nm with a relative uncertainty of about 5%. This series of samples exhibit PMA with coercivity of $\approx 0.4$ T without post-deposition annealing. The saturation magnetization is $M_s = (1.08 \pm 0.05) \times 10^6$ A/m with an apparent "magnetic dead layer" of $t_{\text{FM}}^{\text{dead}} = 0.26 \pm 0.04$ nm [29]. For the HR measurements, the multilayer stacks were patterned into $5 \, \mu\text{m} \times 60 \, \mu\text{m}$ Hall bars by photolithography and ion milling. All measurements were carried out at room temperature (RT).

The sheet conductance of the films were determined by 4-probe resistance measurements of a set of microbars of varying width, length and probe spacing, which minimized errors due to sample geometry and reduced the statistical measurement error to below 1%. Thus the main source of error comes from the uncertainty of film thicknesses. The resistivity of Pt layer $\rho_{Pt}$ was determined by subtracting the sheet conductance of a separately fabricated Ta(1)/Co(1)/MgO(2)/Ta(1) stack from that of our samples containing the Pt layer. In Fig. 1(a) we show $\rho_{Pt}$ for the samples as a function of $t_{Pt}$. The sharp increase of $\rho_{Pt}$ with decreasing $t_{Pt}$ is a well-known phenomenon due to strong diffusive scattering at a Pt surface [48,53–57].

The DL and FL SH torque efficiencies of these PMA samples were measured by the HR technique [20,29] with the same alternating voltage amplitude (4 V) applied to the Hall bars in all measurements, corresponding to an alternating electric field of constant magnitude $E = 67$ kV/m. Fig. 1(b) shows the SH torque-induced longitudinal $H_L$ (corresponding to DL torque) and transverse $H_T$ (corresponding to FL torque) equivalent fields per unit applied electric field determined by the HR measurement as functions of $t_{Pt}$. As $t_{Pt}$ increases, $H_L$ quickly increases and then saturates for $t_{Pt} > 5$ nm. $H_T$ starts for $t_{Pt}$ near zero from a value that is negative in our convention, opposite to the Oersted field generated by the charge current flow in the Pt, but quickly reaches a positive maximum and then decreases gradually. (We will discuss the details of our analysis of this $H_T$ behavior elsewhere.) We determine the DL (FL) SH torque efficiencies *per unit applied current density* as

$$\xi_{DL(FL)}^{j} = \frac{2e}{\hbar}\mu_0 M_s (t_{FM} - t_{FM}^{dead}) \cdot H_{L(T)} / j_e \qquad (1)$$

where $j_e = E/\rho_{Pt}$. Fig. 1(c) shows the DL and FL torque efficiencies per unit current density as functions of $t_{Pt}$. $\xi_{DL}^{j}$ first increases with $t_{Pt}$ and reaches a maximum $\approx 0.12$ at $t_{Pt} = 2.8-3.9$ nm, but then, surprisingly, decreases gradually with $t_{Pt}$. The thickness dependence of $\xi_{DL}^j$ that we observe is qualitatively similar to that observed in YIG/Pt bilayers [57] but quite different from other previous ferromagnetic resonance (FMR) measurements [38,40,44,45] and spin pumping/ISHE experiments [21–23] on metallic FM/Pt bilayers where the data typically are fit by a simple functional form [37]:

$$\xi_{DL}^j(t_{NM}) = \frac{2e}{\hbar} T_{int} j_s(t_{NM}) / j_e(t_{NM}) = \xi_{DL,max}^j \left(1 - \text{sech}(t_{NM}/\lambda_s)\right). \quad (2)$$

This is the behavior expected for an ideal ($T_{int}=1$) interface with no SBF, or alternatively one where SML is the dominant cause for $T_{int} < 1$. However, we emphasize that Eq. (2) holds only under the assumption of constant $\rho_{NM}$ and hence thickness-independent values for $\theta_{SH}$ and $\lambda_s$. In the intrinsic SHE regime, which has recently been reported to describe Pt [41,58], and also in the side-jump regime, it is the spin Hall conductivity $\sigma_{SH}$ that is expected to be constant, independent of $\rho_{NM}(t_{NM})$ while the spin Hall ratio $\theta_{SH}(t_{NM}) = (2e/\hbar)\sigma_{SH}\rho_{NM}(t_{NM})$ should vary $\propto \rho_{NM}(t_{NM})$ and therefore $\xi_{DL}^j$ also depends on the NM resistivity and hence, in this study, on its thickness due to strong interfacial scattering.

An alternative approach is to consider the spin torque efficiency *per unit applied electric field*, determined directly from the HR measurement as

$$\xi_{DL}^E = \frac{2e}{\hbar} \mu_0 M_s (t_{FM} - t_{FM}^{dead}) H_L / E. \quad (3)$$

The dependence of $\xi_{DL}^E$ on Pt thickness is shown in Fig. 1(d) and is consistent with the functional form in Eq. (2) with a prefactor that does not depend on $t_{Pt}$, which indicates that the intrinsic

SHE, or perhaps the side-jump mechanism, is indeed predominant in Pt. Then assuming that (i) the DL torque is entirely due to the SHE of the Pt, (ii) the interface is well ordered, and (iii) SBF is the dominant cause for $T_{int} < 1$, we can expect, approximately, [24,25]

$$\xi_{DL}^E(t_{Pt}) = \frac{2e}{\hbar}\sigma_{SH}\left(1-\text{sech}(t_{Pt}/\lambda_s)\right)\left(1+\frac{\tanh(t_{Pt}/\lambda_s)}{2\lambda_s\rho_{Pt}G_r}\right)^{-1}, \quad (4)$$

where $G_r$ is the real part of the spin mixing conductance $G^{\uparrow\downarrow} = G_r + iG_i$ and we have assumed $G_r \gg G_i$, consistent with our result that $\xi_{DL} \gg \xi_{FL}$. As an exercise, if we fit the $\xi_{DL}^E$ data shown in Fig. 1(d) to equation (4) using a fixed value $\rho_{bulk} = 15\,\mu\Omega\cdot\text{cm}$, the resistivity in the midst of a thick Pt film, and $G_r = 0.59\times10^{15}\,\Omega^{-1}\text{m}^{-2}$ as theoretically calculated for the Pt/Co interface [24], we obtain an "effective" spin diffusion length $\lambda_s^{eff} = 2.0\pm0.1\,\text{nm}$ and $\sigma_{SH} = (10.5\pm0.3)\times10^5\,[\hbar/2e]\,\Omega^{-1}\cdot\text{m}^{-1}$ or $\theta_{SH} = \rho_{bulk}\sigma_{SH} = 0.16\pm0.01$, consistent with previous estimations [28,31]. The choice of $G_r$ may change the estimation of $\sigma_{SH}$ but has a very weak effect on $\lambda_s^{eff}$. The existence of a SML would introduce a constant factor < 1 to the right hand side of equation (4), thus would increase the estimated $\sigma_{SH}$ but would not affect $\lambda_s^{eff}$. (We note that this analysis neglects any possible negative SHE from the 1 nm Ta layer (see the discussion in the Supplementary Material [52]). We account for the maximum possible effect of any SH torque from the Ta underlayer within the experimental uncertainties indicated in Fig. 1(c,d)).

Although $\lambda_s^{eff}$ indicates the scale of the Pt thickness for which the spin current flowing to the FM/NM interface begins to saturate, it is only a phenomenological number since both thickness-independent $\rho_{Pt}$ and $\lambda_s$ are assumed in Eq. (4). In a more realistic approach, given the non-uniformity of resistivity across the layer, both $\theta_{SH}$ and $\lambda_s$ will vary with location within the Pt film. In particular, since the E-Y mechanism [49,50] is expected to be the dominant spin scattering process in Pt, we should have $\lambda_s \propto 1/\rho_{Pt}$. Hence $\lambda_s$ near the Pt interfaces (where $\rho_{Pt}$ is large) should be smaller than in the bulk. This means that the effective $\lambda_s^{eff} = 2.0$ nm obtained above from the simplified equation (4) yields an underestimate of $\lambda_s$ within the bulk of the Pt film.

We have found that it is possible to go beyond this type of approximate treatment and perform, using a simple rescaling, a quantitative calculation of the spin torque (including SBF) even for a heavy-metal layer with a nonuniform resistivity and spin diffusion length, as long as (a) the intrinsic mechanism of the SHE dominates spin current generation and (b) the E-Y mechanism dominates spin relaxation. Assuming that these two conditions hold, we can then use the experimental values of $\xi_{DL}^E(t_{Pt})$ and $\rho_{Pt}(t_{Pt})$ to obtain an estimate for the value of $\lambda_s \rho_{Pt}$.

We first assume, as an exercise, that the thickness-dependence of Pt resistivity is due only to surface scattering at the Pt/Co interface. From the series of $\rho_{Pt}(t_{Pt}^n)$ as a function of Pt thickness presented in Fig. 1(a), we divide each of the Pt films into a series of adjacent "slices" of thickness $l^i$ each of which has a different, but uniform, resistivity $\rho_{Pt}^i$ and spin diffusion length $\lambda_s^i$. These divisions lead to the distribution of $\rho_{Pt}(z)$ as shown in Fig. 2(b), where the z-axis points normal to the layers with $z = 0$ starting at the Pt/Co interface. As fully discussed in the Supplementary Material [52], the spin transmission through the $i$-th slice is identical to that for an "effective" slice having a fixed spin diffusion length $\lambda_s^0$, resistivity $\rho_{Pt}^0$ and a rescaled effective thickness $L^i = l^i \rho_{Pt}^i / \rho_{Pt}^0$ so that $\lambda_s^i \rho_{Pt}^i = \lambda_s^0 \rho_{Pt}^0$ which holds under the E-Y mechanism.

Thus a Pt layer of thickness $t_{Pt}^n = \sum_{i=1}^n l^i$ (a combination of $n$ slices) with non-uniform resistivity and spin diffusion length is equivalent to a uniform "effective" Pt film having a thickness $T_{Pt}^n = \sum_{i=1}^n L^i$, as schematically depicted in Fig. 2(a) for the case of a single interface. The same result can be obtained by the same manner when also including the second interface, i.e., although Fig. 2(a) changes in a way that $\rho_{Pt}(t_{Pt})$ has a minimum somewhere in the midst of the Pt, Fig. 2(b) and the consequent analysis would not change [52]. Since the "effective" layers are chosen to have constant resistivity $\rho_{Pt}^0$ (we choose $15\,\mu\Omega\cdot cm$) and spin diffusion length $\lambda_s^0$, we can fit the $\xi_{DL}^E$ data versus the rescaled thickness $T_{Pt}$ to Eq. (4), just substituting $T_{Pt}$ instead of $t_{Pt}$. One important factor we need to consider is the location of the Pt/Co interface, which is not necessarily at $z = 0$. This is because a few atomic layers of Pt at each of the interfaces may be intermixed with the adjacent material, and/or in the case of the Pt/Co interface magnetized by the proximity effect [59]. This can result in a small offset $t_{off}$ because the thickness of the first slice is smaller than its nominal value. This effect seems to be apparent in Fig. 1(d) where the fitted line (which goes through the origin) does not fit the data particularly well in the thin Pt region. We address this issue in our analysis by replacing $T_{Pt}$ in the right hand side of equation (4) by $T_{Pt} - T_{off}$ where $T_{off}$ is the location of the FM/NM interface and is estimated from the fitting.

The fitted result of the "effective" Pt layers with three free parameters $\sigma_{SH}$, $\lambda_s^0$ and $T_{off}$ is shown in Fig. 2(c). We obtain $\lambda_s^0 = 5.1 \pm 0.5$ nm for $\rho_{Pt}^0 = 15\,\mu\Omega cm$, or more generally we have $\lambda_s \rho_{Pt} = (0.77 \pm 0.08) \times 10^{-15}\,\Omega\cdot m^2$; $T_{off} = 4.9 \pm 0.3$ nm for the "effective" Pt thickness offset which corresponds to $t_{off} = 0.8 \pm 0.1$ nm in the original, un-scaled thickness; and $\sigma_{SH} = (5.9 \pm 0.2) \times 10^5\,[\hbar/2e]\,\Omega^{-1}m^{-1}$ independent of $\rho_{Pt}$, if no SML is present. If we use a somewhat higher $G_r = 1.07 \times 10^{15} \, \Omega^{-1} \mathrm{m}^{-2}$ as calculated including spin orbit effects for the Py/Pt interface [47] then $\sigma_{SH} = (4.5 \pm 0.1) \times 10^5 \, [\hbar/2e] \, \Omega^{-1} \cdot \mathrm{m}^{-1}$, again a lower bound. We reiterate that the existence of SML would increase the estimated $\sigma_{SH}$ but negligibly affect our determination of $\lambda_s^0$. As a final check of this analysis we note the requirement of the E-Y mechanism that the spin relaxation time $\tau_s$ be longer than the momentum scattering time $\tau_m$. It has been reported that the mean free path $l_{mfp}$ in Pt can be estimated from $l_{mfp}[m] \approx 8 \times 10^{-16} / \rho_{Pt} \, [\Omega \cdot \mathrm{m}]$ [60]. Thus we have $\tau_{sf}/\tau_m = 3(\lambda_s/l_{mfp})^2 \approx 3\left[\lambda_s \rho_{Pt}/(8 \times 10^{-16})\right]^2 = 2.8$, which is consistent with the E-Y spin scattering mechanism being dominant in Pt.

We now discuss our results in relation with previous results in the literature. First, as noted above, previous ST-FMR and ISHE studies on in-plane magnetized (IPM) Pt/Py bilayers [38,40,44] did not yield a peak in the apparent damping-like spin torque efficiency as a function of $t_{Pt}$ such as reported here. These previous analyses also reported a short $\lambda_s \approx 1.4$ nm as determined by RT ST-FMR, or alternatively by ISHE, on Py/Pt [38,40,44] and $\lambda_s \approx 2.1$ nm for Co$_{75}$Fe$_{25}$/Pt [45], in the same range as $\lambda_s^{eff} = 2.0$ nm. These differences with our results can be explained by a weaker thickness dependence of the resistivity for multilayers made from different materials and the neglect of any field-like torque in the analysis. See the Supplementary Material [52] for further discussion on these points.

An alternative approach to estimate $\lambda_s$ is to measure the $t_{Pt}$ dependence of Gilbert magnetic damping in bilayer samples, and such a study has recently reported $\lambda_s = 0.5 \pm 0.3$ nm [42]. Fast saturation of the damping at very thin Pt thicknesses has also been observed previously [22,37,38]. However, Liu *et al.* [47] have recently pointed out that this very rapid attenuation is likely due to strong SML at the FM/Pt interface, and used a first principles calculation and data [61] from this measurement method to obtain $\lambda_s \approx 5.5$ nm, or more generally $\lambda_s \rho_{Pt} = (0.61 \pm 0.02) \times 10^{-15}$ $\Omega \cdot m^2$. On the other hand, a longer $\lambda_s \approx 8.0$ nm has been reported [21,22] from ISHE experiments on Py/Pt at RT. However, these latter works did not consider SML or spin backflow at the FM-NM interface which would reduce their estimated values, as pointed out by Jiao *et al.* [43]. Rojas-Sanchez *et al.* [23] performed similar measurement on Co/Pt and, after taking SML into account, reported $\lambda_s = 3.4 \pm 0.4$ nm and $\lambda_s \rho_{Pt} = (0.59 \pm 0.06) \times 10^{-15}$ $\Omega \cdot m^2$. These experiments did not consider the non-uniformity of the local resistivity $\rho_{Pt}(t_{Pt})$ and its effect on $\lambda_s(t_{Pt})$, and thus underestimated the value of $\lambda_s \rho_{Pt}$. A very high value $\lambda_s = 11 \pm 2$ nm has been determined from a low temperature, 3-10 K, study of spin pumping in lateral spin valves [39,41] for samples having $\rho_{Pt} = 12$ $\mu\Omega \cdot cm$, or $\lambda_s \rho_{Pt} = 1.32 \times 10^{-15}$ $\Omega \cdot m^2$. However, Isasa *et al.* used a similar lateral spin value technique and reported $\lambda_s \rho_{Pt} = (0.85 \pm 0.08) \times 10^{-15}$ $\Omega \cdot m^2$ at 10 K and $(0.79 \pm 0.87) \times 10^{-15}$ $\Omega \cdot m^2$ at RT [58], while measurements using current-perpendicular to the plane studies of Py-based exchange biased spin valves [26] at 4.2 K have reported $\lambda_s \rho_{Pt} = (0.59 \pm 0.25) \times 10^{-15}$ $\Omega \cdot m^2$ [37] and $(0.72 \pm 0.13) \times 10^{-15}$ $\Omega \cdot m^2$ [46]. All of these latter results are in reasonable agreement with our result $\lambda_s \rho_{Pt} = (0.77 \pm 0.08) \times 10^{-15}$ $\Omega \cdot m^2$.

In summary, we have observed a strong dependence on $t_{Pt}$ for the damping-like SH torque efficiency per unit applied current density for perpendicularly-magnetized Pt/Co bilayer structures, with a peak value $\xi_{DL}^j = 0.12$ at $t_{Pt} = 2.8 - 3.9$ nm, while the spin torque efficiency per unit applied electric field exhibits a monotonic increase with increasing Pt thickness and saturates for $t_{Pt} > 5$ nm. We interpret this behavior as an indication that the intrinsic SHE being the dominant SHE mechanism in Pt, perhaps in combination with side-jump scattering, so that the SH conductivity is independent of mean free path while the SH torque efficiency per unit current density is enhanced by an increased $\rho_{Pt}(t_{Pt})$ associated with interfacial scattering. By assuming the E-Y mechanism for spin scattering, which implies that $\lambda_s \propto 1/\rho_{Pt}$ so that $\lambda_s$ is also non-uniform, we obtain $\lambda_s \rho_{Pt} = (0.77 \pm 0.08) \times 10^{-15}\ \Omega \cdot m^2$. With this result we can apply SBF analysis to our direct measurements of $\xi_{DL}^E$ for this PMA system using $G_r = 0.59 \times 10^{15}\ \Omega^{-1} m^{-2}$ [24], and obtain $\sigma_{SH}^{Pt} = (5.9 \pm 0.2) \times 10^5\ [\hbar/2e]\ \Omega^{-1} \cdot m^{-1}$, with this being a lower bound as it is made with the assumption that there is no significant SML at our Pt/Co interfaces.

This work seems to resolve the controversy regarding the differences in the value of $\lambda_s$ for Pt as obtained from various spin Hall and other experiments, and demonstrates that the spin Hall efficiency of Pt can be enhanced by increasing its resistivity, as expected when the intrinsic SHE is dominant.

**Acknowledgments**

We thank Y. Ou, C.-F. Pai and S. Emori for fruitful discussions, G. E. Rowlands for technical support and F. Guo for commenting on the manuscript. This work was supported in part by the Samsung Electronics Corporation, by the NSF/MRSEC program (DMR-1120296) through, the Cornell Center for Materials Research, and by ONR. We also acknowledge support from the NSF (Grant No. ECCS-0335765) through use of the Cornell Nanofabrication Facility/National Nanofabrication Infrastructure Network.

# REFERENCES

[1] M. I. Dyakonov and V. I. Perel, Phys. Leters **35A**, (1971).

[2] J. Hirsch, Phys. Rev. Lett. **83**, 1834 (1999).

[3] S. Zhang, Phys. Rev. Lett. **85**, 393 (2000).

[4] C.-F. Pai, L. Liu, Y. Li, H. W. Tseng, D. C. Ralph, and R. A. Buhrman, Appl. Phys. Lett. **101**, 122404 (2012).

[5] L. Liu, C.-F. Pai, D. C. Ralph, and R. A. Buhrman, Phys. Rev. Lett. **109**, 186602 (2012).

[6] V. E. Demidov, S. Urazhdin, H. Ulrichs, V. Tiberkevich, A. Slavin, D. Baither, G. Schmitz, and S. O. Demokritov, Nat. Mater. **11**, 1028 (2012).

[7] V. E. Demidov, H. Ulrichs, S. V Gurevich, S. O. Demokritov, V. S. Tiberkevich, a N. Slavin, a Zholud, and S. Urazhdin, Nat. Commun. **5**, 3179 (2014).

[8] T. Jungwirth, J. Wunderlich, and K. Olejník, Nat. Mater. **11**, 382 (2012).

[9] P. P. J. Haazen, E. Murè, J. H. Franken, R. Lavrijsen, H. J. M. Swagten, and B. Koopmans, Nat. Mater. **12**, 299 (2013).

[10] N. Okamoto, H. Kurebayashi, T. Trypiniotis, I. Farrer, D. A. Ritchie, E. Saitoh, J. Sinova, J. Mašek, T. Jungwirth, and C. H. W. Barnes, Nat. Mater. **13**, 932 (2014).

[11] D. Bhowmik, L. You, and S. Salahuddin, Nat. Nanotechnol. **9**, 59 (2013).

[12] D. M. Bromberg, D. H. Morris, L. Pileggi, and J.-G. Zhu, IEEE Trans. Magn. **48**, 3215 (2012).

[13] S. Urazhdin, V. E. Demidov, H. Ulrichs, T. Kendziorczyk, T. Kuhn, J. Leuthold, G. Wilde, and S. O. Demokritov, Nat. Nanotechnol. **9**, 509 (2014).

[14] G. Vignale, J. Supercond. Nov. Magn. **23**, 3 (2010).

[15] A. Hoffmann, IEEE Trans. Magn. **49**, 5172 (2013).

[16] R. Karplus and J. M. Luttinger, Phys. Rev. **95**, 1154 (1954).

[17] L. Berger, Phys. Rev. B **2**, 4559 (1970).

[18] J. Smit, Physica **24**, 39 (1958).

[19] L. Liu, C.-F. Pai, Y. Li, H. W. Tseng, D. C. Ralph, and R. A. Buhrman, Science **336**, 555 (2012).

[20] J. Kim, J. Sinha, M. Hayashi, M. Yamanouchi, S. Fukami, T. Suzuki, S. Mitani, and H. Ohno, Nat. Mater. **12**, 240 (2013).

[21] H. Nakayama, K. Ando, K. Harii, T. Yoshino, R. Takahashi, Y. Kajiwara, K. Uchida, Y. Fujikawa, and E. Saitoh, Phys. Rev. B **85**, 144408 (2012).

[22] Z. Feng, J. Hu, L. Sun, B. You, D. Wu, J. Du, W. Zhang, A. Hu, Y. Yang, D. M. Tang, B. S. Zhang, and H. F. Ding, Phys. Rev. B **85**, 214423 (2012).

[23] J.-C. Rojas-Sánchez, N. Reyren, P. Laczkowski, W. Savero, J.-P. Attané, C. Deranlot, M. Jamet, J.-M. George, L. Vila, and H. Jaffrès, Phys. Rev. Lett. **112**, 106602 (2014).

[24] P. M. Haney, H.-W. Lee, K.-J. Lee, A. Manchon, and M. D. Stiles, Phys. Rev. B **87**, 174411 (2013).

[25] Y.-T. Chen, S. Takahashi, H. Nakayama, M. Althammer, S. Goennenwein, E. Saitoh, and G. Bauer, Phys. Rev. B **87**, 144411 (2013).

[26] W. Park, D. Baxter, S. Steenwyk, I. Moraru, W. Pratt, and J. Bass, Phys. Rev. B **62**, 1178 (2000).

[27] L. Liu, T. Moriyama, D. C. Ralph, and R. A. Buhrman, Phys. Rev. Lett. **106**, 036601 (2011).

[28] M.-H. Nguyen, C.-F. Pai, K. X. Nguyen, D. A. Muller, D. C. Ralph, and R. A. Buhrman, Appl. Phys. Lett. **106**, 222402 (2015).

[29] C.-F. Pai, Y. Ou, L. H. Vilela-leão, D. C. Ralph, and R. A. Buhrman, Phys. Rev. B **92**, 064426 (2015).

[30] L. Liu, C.-T. Chen, and J. Z. Sun, Nat. Phys. **10**, 561 (2014).

[31] W. Zhang, W. Han, X. Jiang, S.-H. Yang, and S. S. P. Parkin, Nat. Phys. **11**, 496 (2015).

[32] X. Fan, J. Wu, Y. Chen, M. J. Jerry, H. Zhang, and J. Q. Xiao, Nat. Commun. **4**, 1799 (2013).

[33] X. Fan, H. Celik, J. Wu, C. Ni, K.-J. Lee, V. O. Lorenz, and J. Q. Xiao, Nat. Commun. **5**, 3042 (2014).

[34] T. Nan, S. Emori, C. T. Boone, X. Wang, T. M. Oxholm, J. G. Jones, B. M. Howe, G. J. Brown, and N. X. Sun, Phys. Rev. B **91**, 214416 (2015).

[35] T. D. Skinner, M. Wang, A. T. Hindmarch, A. W. Rushforth, A. C. Irvine, D. Heiss, H. Kurebayashi, and A. J. Ferguson, Appl. Phys. Lett. **104**, 062401 (2014).

[36] K. Garello, I. M. Miron, C. O. Avci, F. Freimuth, Y. Mokrousov, S. Blügel, S. Auffret, O. Boulle, G. Gaudin, and P. Gambardella, Nat. Nanotechnol. **8**, 587 (2013).

[37] H. Kurt, R. Loloee, K. Eid, W. P. Pratt, and J. Bass, Appl. Phys. Lett. **81**, 4787 (2002).

[38] L. Liu, R. A. Buhrman, and D. C. Ralph, arXiv:**1111**.3702 (2011).

[39] M. Morota, Y. Niimi, K. Ohnishi, D. H. Wei, T. Tanaka, H. Kontani, T. Kimura, and Y. Otani, Phys. Rev. B **83**, 174405 (2011).

[40] K. Kondou, H. Sukegawa, S. Mitani, K. Tsukagoshi, and S. Kasai, Appl. Phys. Express **5**, 073002 (2012).

[41] Y. Niimi, D. Wei, H. Idzuchi, T. Wakamura, T. Kato, and Y. Otani, Phys. Rev. Lett. **110**, 016805 (2013).

[42] C. T. Boone, H. T. Nembach, J. M. Shaw, and T. J. Silva, J. Appl. Phys. **113**, 153906 (2013).

[43] H. Jiao and G. E. W. Bauer, Phys. Rev. Lett. **110**, 217602 (2013).

[44] W. Zhang, V. Vlaminck, J. E. Pearson, R. Divan, S. D. Bader, and A. Hoffmann, Appl. Phys. Lett. **103**, 242414 (2013).

[45] A. Ganguly, K. Kondou, H. Sukegawa, S. Mitani, S. Kasai, Y. Niimi, Y. Otani, and A. Barman, Appl. Phys. Lett. **104**, 072405 (2014).

[46] H. Y. T. Nguyen, W. P. Pratt, and J. Bass, J. Magn. Magn. Mater. **361**, 30 (2014).

[47] Y. Liu, Z. Yuan, R. J. H. Wesselink, A. A. Starikov, and P. J. Kelly, Phys. Rev. Lett. **113**, 207202 (2014).

[48] C. T. Boone, J. M. Shaw, H. T. Nembach, and T. J. Silva, J. Appl. Phys. **117**, 223910 (2015).

[49] R. J. Elliott, Phys. Rev. **96**, 266 (1954).

[50] Y. Yafet, Solid State Phys. **14**, 1 (1963).

[51] J. M. Shaw, H. T. Nembach, T. J. Silva, S. E. Russek, R. Geiss, C. Jones, N. Clark, T. Leo, and D. J. Smith, Phys. Rev. B **80**, 184419 (2009).

[52] See Supplementary at [URL] for further discussion on the rescaling method, the previously reported ST-FMR studies and the effect of the Ta seeding layer.

[53] A. F. Mayadas and M. Shatzkes, Phys. Rev. B **1**, 1382 (1970).

[54] F. Warkusz, Electrocompon. Sci. Technol. **5**, 99 (1978).

[55] H. D. Liu, Y. P. Zhao, G. Ramanath, S. P. Murarka, and G. C. Wang, Thin Solid Films **384**, 151 (2001).

[56] W. E. Bailey, S. X. Wang, and E. Y. Tsymbal, Phys. Rev. B **61**, 1330 (2000).

[57] V. Castel, N. Vlietstra, J. Ben Youssef, and B. J. Van Wees, Appl. Phys. Lett. **101**, 132414 (2012).

[58] M. Isasa, E. Villamor, L. E. Hueso, M. Gradhand, and F. Casanova, Phys. Rev. B **91**, 024402 (2014).

[59] S. Y. Huang, X. Fan, D. Qu, Y. P. Chen, W. G. Wang, J. Wu, T. Y. Chen, J. Q. Xiao, and C. L. Chien, Phys. Rev. Lett. **109**, 107204 (2012).

[60] J. Bass and W. P. Pratt, J. Phys. Condens. Matter **19**, 183201 (2007).

[61] S. Mizukami, Y. Ando, and T. Miyazaki, Phys. Rev. B **66**, 104413 (2002).

**FIGURES**

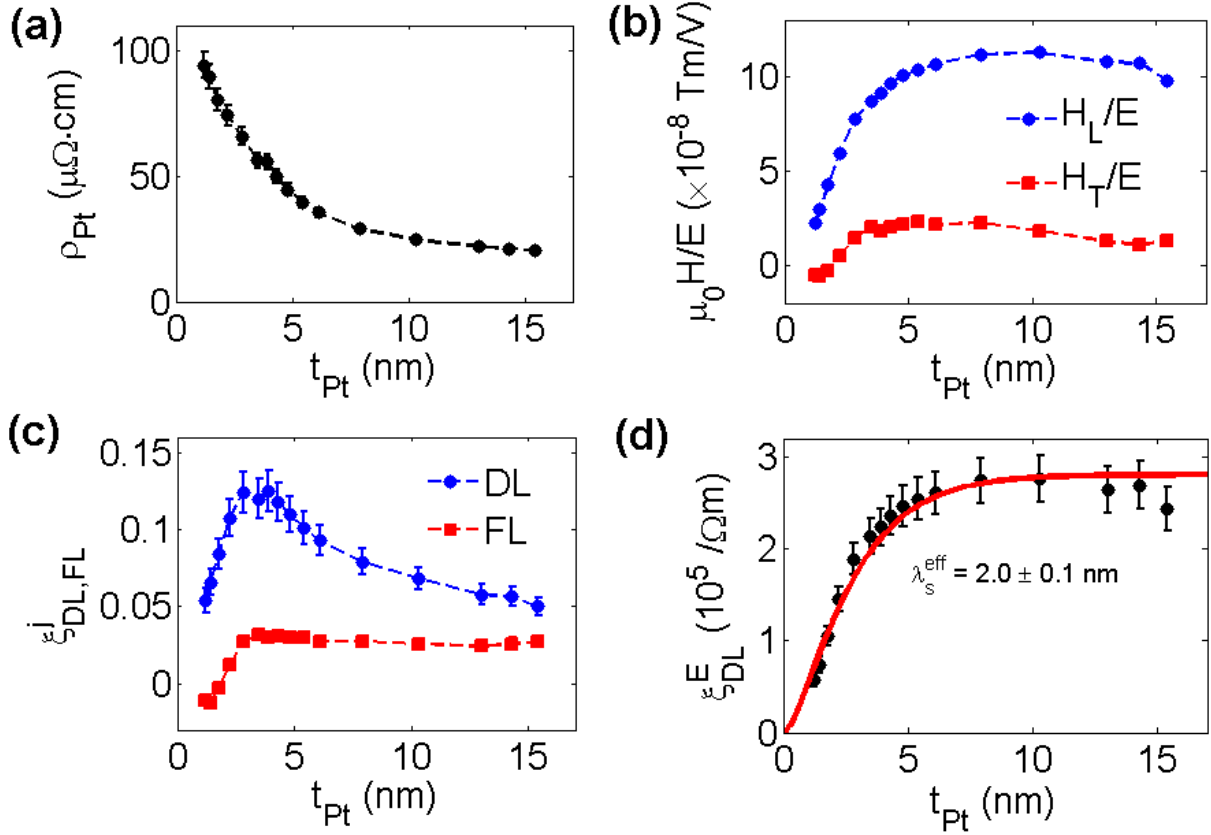

**Figure 1:** (Color online) **(a)** Resistivity of Pt in Ta(1)/Pt/Co(1), **(b)** SH torque-induced longitudinal (circles) and transverse (squares) equivalent fields per unit applied electric field, **(c)** damping-like (circles) and field-like (squares) SH torque efficiency per unit applied current density, and **(d)** damping-like SH torque efficiency per unit applied electric field as functions of Pt thickness. The solid line in (d) shows the fitted result to equation (4) from which the *effective* spin diffusion length is estimated to be $\lambda_s^{eff} = 2.0 \pm 0.1\,\text{nm}$. The broken lines in other plots connect the data points.

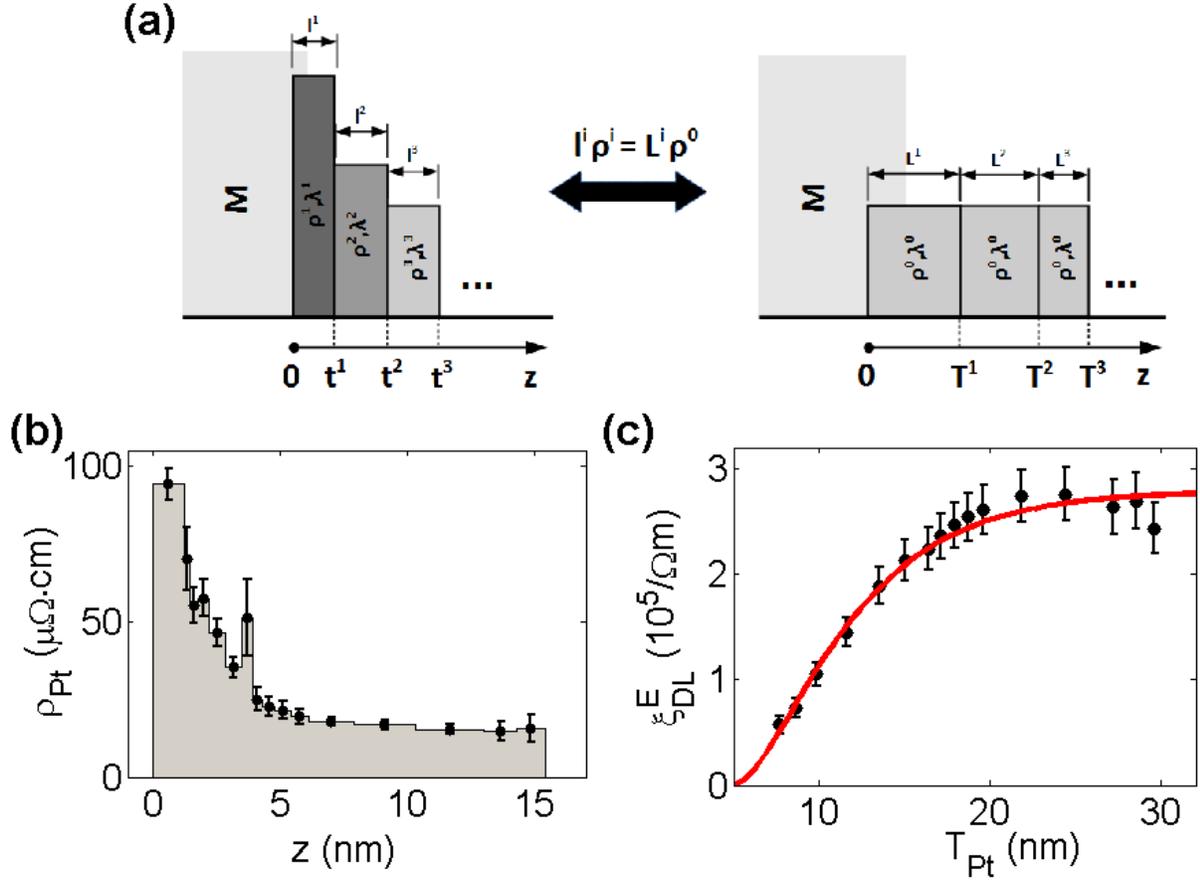

**Figure 2:** (Color online) Estimation of spin diffusion length within the E-Y mechanism. **(a)** Schematic illustration of the "slicing" and "rescaling" process in which a non-uniform layer $t_{Pt}^n$ is scaled into a uniform one $T_{Pt}^n$. See full description in the main text. **(b)** The distribution of Pt local resistivity with location $z$, extracted from the experimental Ta/Pt/Co data in Fig. 1(a). The points represent the local resistivity of each "slice". **(c)** Damping-like spin torque efficiency per unit applied electric field versus "effective" thickness $T_{Pt}$. The solid line shows the fitted result from which the spin diffusion length of Pt at $\rho_{Pt}^0 = 15\ \mu\Omega\text{cm}$ is estimated to be $\lambda_s^0 = 5.1 \pm 0.5$ nm.

# SUPPLEMENTARY MATERIAL

# Spin torque study of the spin Hall conductivity and spin diffusion length in platinum thin films

Minh-Hai Nguyen[1], D. C. Ralph[1,2], R. A. Buhrman[1*]

[1]Cornell University, Ithaca, New York 14853, USA

[2]Kavli Institute at Cornell, Ithaca, New York 14853, USA

Contents

---

* rab8@cornell.edu

1. **Discussion of the details of the rescaling analysis**

   *a)* *For one interface*

   In the main text, we treat the non-uniform $\rho_{Pt}(t_{Pt}^n)$ by dividing each the Pt layer with thickness $t_{Pt}^n$ into $n$ slices of thickness $l^i$ and uniform resistivity $\rho_{Pt}^i$ so that

   $$\frac{t_{Pt}^n}{\rho_{Pt}(t_{Pt}^n)} = \sum_{i=1}^{n} \frac{l^i}{\rho_{Pt}^i}. \qquad (S1)$$

   For the case of one interface between Pt and another metal layer (M), the process is illustrated in Fig. 2(a)-(b) in the main text. To account for the variation of $\lambda_s^i$ from slice to slice, let us consider the transmission of spin current through the *i*-th layer. For a simple drift-diffusion model [1], let $\mu_s(z)$ be the spin potential satisfying the diffusion equation $\nabla^2 \mu_s(z) = \mu_s(z)/(\lambda_s^i)^2$, and $j_s(z) = j_{SHE} - \nabla \mu_s(z)/\rho^i$ be the local spin current, where $j_{SHE} = \sigma_{SH} \cdot E$ is the SHE-induced spin current which is constant, the same for all slices, since the applied electric field is the same in all slices and $s_{SH}$ is independent of resistivity if the intrinsic SHE mechanism dominates. Elementary calculus shows that the relation of $\{j_s(z), \mu_s(z)\}$ at the two surfaces $z = z_1, z_2$ of the *i*-th slice can be expressed as [2]

   $$\begin{pmatrix} j_s(z_1) - j_{SHE} \\ \mu_s(z_1) \end{pmatrix} = \Pi(l^i/\lambda_s^i, \lambda_s^i \rho^i) \cdot \begin{pmatrix} j_s(z_2) - j_{SHE} \\ \mu_s(z_2) \end{pmatrix}, \qquad (S2)$$

   where

   $$\Pi(\alpha, \beta) \equiv \begin{pmatrix} \cosh \alpha & (\sinh \alpha)/\beta \\ (\sinh \alpha) \cdot \beta & \cosh \alpha \end{pmatrix} \qquad (S3)$$

is the spin transmission matrix of the $i$-th slice. Since $\Pi(l^i/\lambda_s^i, \lambda_s^i \rho^i)$ depends only on $\lambda_s^i \rho_{Pt}^i$ =constant (E-Y mechanism) and $l^i/\lambda_s^i$, the calculation for the $i$-th slice is identical to that for an "effective" slice having a fixed spin diffusion length $\lambda_s^0$, resistivity $\rho_{Pt}^0$ and a rescaled effective thickness $L^i = l^i \rho_{Pt}^i / \rho_{Pt}^0$ so that $\lambda_s^i \rho_{Pt}^i = \lambda_s^0 \rho_{Pt}^0$ which holds under the E-Y mechanism. Thus a Pt layer of thickness $t_{Pt}^n = \sum_{i=1}^{n} l^i$ (a combination of $n$ slices) with non-uniform resistivity and spin diffusion length is equivalent to a uniform "effective" Pt film having a thickness

$$T_{Pt}^n = \sum_{i=1}^{n} L^i = \sum_{i=1}^{n} l^i \rho_{Pt}^i / \rho_{Pt}^0. \quad (S4)$$

The analysis using this scaling process performed in the main text assumed the surface scattering occurred at only the Pt/Co interface. If the surface scattering at the Ta/Pt interface is a major contributor to the increase of $\rho_{Pt}$ in the thin Pt region the slicing process will result in a distribution of $\rho_{Pt}$ that is horizontally opposite to Fig. 2(b). Since the spin transmission matrix $\Pi(l/\lambda_s, \lambda_s \rho)$ defined by Eq. (S3) is *commutative within the E-Y mechanism*, i.e. if $\lambda_s^1 \rho^1 = \lambda_s^2 \rho^2$ then

$$\Pi(l^1/\lambda_s^1, \lambda_s^1 \rho^1) \cdot \Pi(l^2/\lambda_s^2, \lambda_s^2 \rho^2) = \Pi(l^2/\lambda_s^2, \lambda_s^2 \rho^2) \cdot \Pi(l^1/\lambda_s^1, \lambda_s^1 \rho^1), \quad (S5)$$

we can interchange any two adjacent slices without affecting the spin transmission through the two slices. As a result, the slices can be rearranged in an arbitrary order and therefore the same result will be obtained whether the interfacial scattering occurs predominately at either the Ta/Pt or the Pt/Co interface.

*b)*    *For two interfaces*

We now prove that the rescaling process would yield the same result for the case where the surface scattering occurs at two interfaces. The slicing process still follows relation (S1) but the location of the *(i+1)*-th slice is not at the far end of the *i*-th slice as in the case of one interface, instead it is somewhere in the midst of the *i*-th slice. To model this difference, we notice that

$$\Pi((l_1+l_2)/\lambda_s, \lambda_s \rho) = \Pi(l_1/\lambda_s, \lambda_s \rho) \cdot \Pi(l_2/\lambda_s, \lambda_s \rho) \qquad (S6)$$

which allows us to further divide the *i*-th slice into two thinner slices with thickness $l_L^i$ and $l_R^i$ (*not* necessarily the same) so that $l^i = l_L^i + l_R^i$, between which the *(i+1)*-th slice is located. The "slicing" process is carried out in the same manner for the next slices and illustrated in Fig. S1(a)-(c). The resulted distribution of $\rho_{Pt}$ with location is now different from the case of one interface in that $\rho_{Pt}$ is minimum somewhere in the midst of Pt layer. However, after rescaling the slices by the same rule $L_{L(R)}^i = l_{L(R)}^i \rho_{Pt}^i / \rho_{Pt}^0$, as illustrated in Fig. S1(d), the original Pt layer of thickness $t_{Pt}^n$ and non-uniform resistivity becomes a uniform layer of fixed $\rho_{Pt}^0$ and thickness

$$T_{Pt}^n = \sum_{i=1}^{n}(L_L^i + L_R^i) = \sum_{i=1}^{n}(l_L^i + l_R^i)\rho_{Pt}^i / \rho_{Pt}^0 = \sum_{i=1}^{n} l^i \rho_{Pt}^i / \rho_{Pt}^0 \qquad (S7)$$

which is the same as (S4). Therefore, the rescaling process yields the same final result for both cases of one (either Ta/Pt or Pt/Co) and two interfaces.

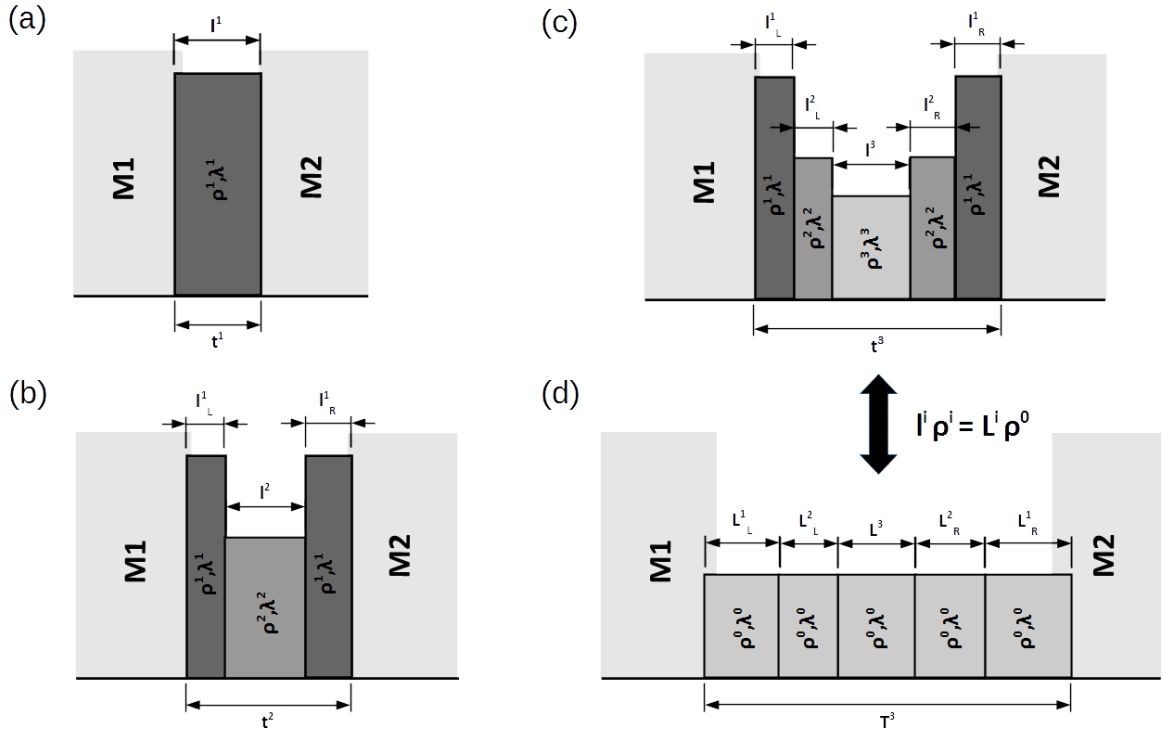

**Figure S1.** Illustration of the scaling process for two interfaces: **(a)-(c)** Illustration of the "slicing" process in which the *(i+1)*-th slice is placed in the midst of the *i*-th slice. **(d)** Result of the rescaling process (shown for *n* = 3).

## 2. Discussion of previous SHE measurements on Py/Pt bilayers

Here we discuss the previously reported determinations of spin diffusion length $\lambda_s$ using spin torque ferromagnetic resonance (ST-FMR) measurement [3–5] and inverse spin Hall effect (ISHE) [6] with changing thickness of Pt layer. First, we note that previous ST-FMR and ISHE studies on in-plane magnetized (IPM) Pt/Py bilayers [3,4,6] did not yield a peak in the apparent damping-like spin torque efficiency $\xi_{FMR}^j$ as a function of $t_{Pt}$ such as reported here. Second, those early works estimated a short estimated spin diffusion length $\lambda_s \approx 1.4$ nm. These discrepancies with our results can be attributed to two main causes:

i. The neglecting of any possible field-like torque in the analysis.

ii. The weaker thickness dependence of the average electrical resistivity of the Pt layers.

### a) *Effect of a field-like torque in ST-FMR measurement*

In many ST-FMR measurements of NM/FM bilayers, a torque efficiency $\xi_{FMR}^j$ is determined by the ratio of the symmetric and anti-symmetric components of the FMR lineshape. If no significant field-like spin transfer torque effect is present, the anti-symmetric component is due to only the Oersted field from the electric current flowing in the NM layer. In that ideal case, $\xi_{FMR}^j = \xi_{DL}^j$. However, if a significant field-like torque $\xi_{FL}$ is present, the anti-symmetric component of ST-FMR lineshape is the combined effect of both the Oersted field and $\xi_{FL}$. Thus in a more general case, $\xi_{FMR}^j$ can be expressed as [7]

$$\frac{1}{\xi_{FMR}^{j}} = \frac{1}{\xi_{DL}^{j}}\left(1 + \frac{\hbar}{e}\frac{\xi_{FL}^{j}}{\mu_0 M_s t_{FM}^{eff} t_{NM}}\right) \qquad (S8)$$

where $t_{FM}^{eff}$ is the effective thickness of the FM layer and $t_{NM}$ is the NM thickness. Of course, both $\xi_{DL}^{j}$ and $\xi_{FL}^{j}$ can be thickness dependent, as demonstrated in the main text. Therefore the determination of $\xi_{DL}^{j}$ from $\xi_{FMR}^{j}$ is not straightforward.

To illustrate the effect of field-like torque on the thickness dependence of $\xi_{FMR}^{j}$, we attempt to estimate $\xi_{DL}^{j}$ from the values of $\xi_{FMR}^{j}$ in Ref. [3], using the value $\xi_{FL}^{j} = +0.024$ which was recently estimated for Py(2.5)/Pt(4) (Py = Ni$_{80}$Fe$_{20}$, thickness in nanometer) [8]. For the purpose of this discussion we assume $\xi_{FL}^{j}$ to be independent of $t_{Pt}$. Using equation (S8), $\xi_{DL}^{j}$ is estimated and is shown in Fig. S2(a) as a function of $t_{Pt}$. A peak in $\xi_{DL}^{j}$ is clearly seen about $t_{Pt} = 2-3$ nm, which is similar in location to the result (Fig. 1 (c)) described in the main text, but less pronounced in amplitude due to the lower variation of $r_{Pt}(t_{Pt})$ in the Pt/Py bilayers.

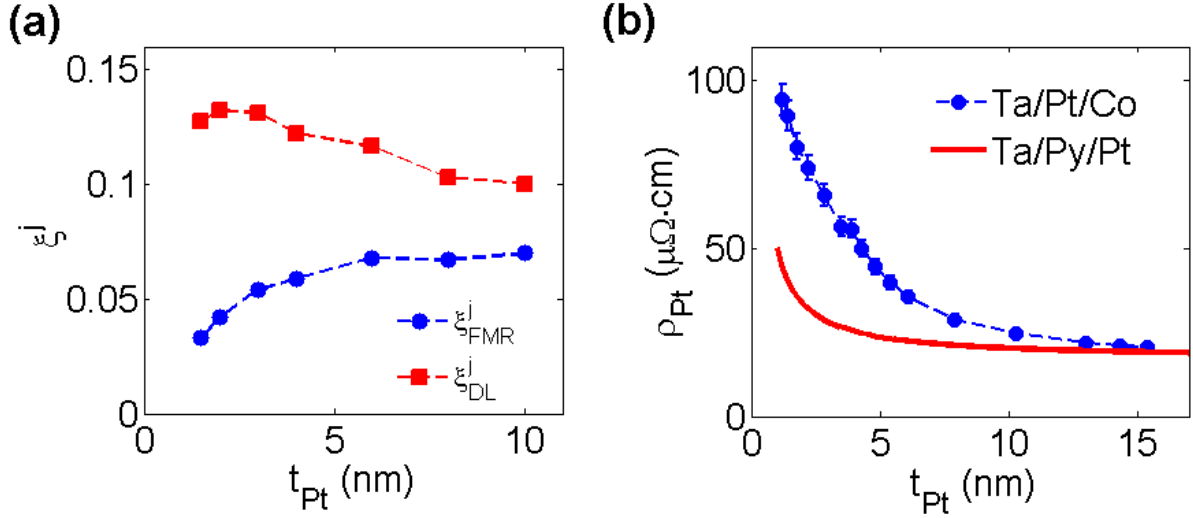

**Figure S2**. (a) $\xi_{FMR}^j$ determined from ST-FMR measurements on Py/Pt, taken from Ref. [3], and the estimated $\xi_{DL}^j$ using equation (S8) with $\xi_{FL}^j = +0.024$ reported by Ref. [8]. A small peak in $\xi_{DL}^j$ is seen about $t_{Pt} = 2-3$ nm. (b) Pt resistivity in *substrate*/Ta(1)/Pt/Co(1) multilayers (points) and *substrate*/Ta(3)/Py(3)/Pt multilayers (solid line, estimated in Ref. [2]) as functions of $t_{Pt}$. The dashed lines connect the data points.

*b)*   *Thickness dependence of resistivity*

Most experimental studies on the SHE and ISHE so far have assumed a uniform electrical resistivity $\rho$, thus a constant spin Hall ratio. In a more realistic situation, the resistivity is non-uniform due to surface scattering when the film is very thin, comparable to its mean free path. As discussed in the main text, in the intrinsic spin Hall effect (SHE), which was shown to be dominant in Pt, the spin Hall ratio is proportional to the resistivity. In multilayers having strong surface scattering at Pt interfaces, $\xi_{DL}^{j}$ depends on both the relative magnitude of the Pt thickness to its spin diffusion length *and* the electrical resistivity $\rho_{Pt}$, and thus may exhibit a peak as in our study. However in the earlier bilayer studies [3–6] only one surface of the Pt layer was adjacent to a metal, a rather thick and relatively low resistivity Py layer, while the other surface was either an oxide (e.g., $SiO_2$ or $AlO_x$) or simply exposed to air. These differences substantially reduce the diffusive surface scattering contributions to the Pt resistivity compared to that which occurs in our *substrate*/Ta(1)/Pt/Co(1) trilayer structures, and thus while $\rho_{Pt}^{Py/Pt}(t_{Pt})$ still varies with $t_{Pt}$ in the thin limit, the variation is less strong than for $\rho_{Pt}^{Ta/Pt/Co}(t_{Pt})$.

Fig. S2(b) shows $\rho_{Pt}$ for our *substrate*/Ta(1)/Pt/Co(1) samples (points) and for *substrate*/Ta(3)/Py(3)/Pt structure (solid line) which was estimated in Ref. [2]. It is clearly seen that $\rho_{Pt}$ in Ta/Pt/Co has a stronger $t_{Pt}$ dependence, which is due to stronger surface scattering with Ta and Co layers than Py/Pt. This weaker thickness dependence of $\rho_{Pt}$ in Py/Pt bilayer contributes to the absence of a peak in $\xi_{DL}^{j}$ as reported by Ref. [3–5].

Nevertheless, as long as there is increased scattering at the Pt/FM interface and the E-Y spin scattering mechanism is dominant, then any analysis which assumes a constant $\lambda_s(t_{Pt})$ will result in an underestimate of its "bulk" value. This explains the short $\lambda_s \approx 1.4\,\text{nm}$ as determined by RT ST-FMR on Py/Pt [3,4,6] and $\lambda_s \approx 2.1\,\text{nm}$ for $Co_{75}Fe_{25}$/Pt [5], in the same range as $\lambda_s^{\text{eff}} = 2.0 \pm 0.1\,\text{nm}$ that we obtained by assuming a constant $\lambda_s$ in fitting to Eq. (4).

### 3. Effect of 1 nm Ta seeding layer

The multilayers in our study were grown on oxidized Si substrates with 1 nm Ta seeding layer. Due to the strong bonding of Ta ions to oxide surfaces and to the strong metallic bonding that occurs between Ta and most transition metals, a thin Ta seeding layer is known to be effective for serving as a strong adhesion and smoothing layer, reducing the grain size of the upper layer, as discussed in details in Ref. [9], and is widely used in spintronics studies [2,7,8] and applications. To minimize the spin torque contributions from the Ta seeding layer, its thickness was chosen to be low, 1 nm. The resistivity of the 1 nm Ta seed layer was $560\,\mu\Omega\cdot\mathrm{cm}$ as determined by a 4-probe resistance measurement of a multilayer stack consisting of substrate/Ta(1)/MgO(1)/Ta(1, oxidized cap). To examine the effect of the 1 nm Ta seeding layer in our multilayers, we measured the averaged resistivity of Pt thin films sputtered directly on the substrate and compared to films deposited on a 1 nm Ta seeding layer (after subtracting of the contribution from the 1 nm Ta layer). The results are shown on Fig. S3 along with the averaged resistivity of Pt in the multilayers in our study (which is shown on Fig. 1(a) in the main text). It is clear from Fig. S3 that the Pt film when deposited onto a bare oxide substrate (red) is very resistive when its average thickness is less than 2.5 nm, with the resistivity rising very quickly with decreasing thickness below this point, while for thicker Pt films on $SiO_2$ the resistivity is slightly less than when deposited on Ta. This behavior is directly attributable a comparably large grain size in the Pt films deposited on oxide that is due to the lack of strong adhesion between the Pt atoms and the oxide surface, which leads to less than full coverage of the surface in the less than 2.5 nm thickness range. The resistivity of Pt film deposited on a 1 nm Ta seeding layer (blue) is much lower in the thin regime, indicating the role of the Ta seeding layer in smoothing the surface and reducing Pt grain size. Finally, the Pt resistivity in the Ta/Pt/Co multilayer used in our study is very similar to that of Ta/Pt sample, indicating that the diffusive scattering at Ta/Pt interface is dominant in increasing Pt resistivity in thin Pt regime.

Next, we estimate the contribution to the spin torques in the Co layer from the 1 nm Ta seeding layer. It has been shown that high resistivity beta-phase Ta has a negative spin Hall ratio [10] whose magnitude is about 0.11 - 0.15. The strength of the contribution of the current in the Ta layer to the spin torque on the Co will depend on whether the SHE in Ta is intrinsic or extrinsic:

**Case 1:** If the dominant mechanism of SHE in Ta is intrinsic and/or side-jump then the spin Hall conductivity of Ta is constant, independent of Ta thickness and resistivity. Taking the values reported in Ref. [10], we can estimate

$$\sigma_{SH}^{Ta} = (\hbar/2e) \cdot \theta_{SH}^{Ta} / \rho_{Ta} \approx -0.79 \times 10^5 \, [\hbar/2e] \, \Omega^{-1} \cdot m^{-1} \quad (S9)$$

**Case 2:** If the dominant mechanism for SHE in Ta is skew-scattering then the spin Hall ratio $\theta_{SH}^{Ta}$ is constant, independent of Ta thickness and resistivity. Since the measured resistivity of our 1 nm Ta seeding layer $\approx 560 \, \mu\Omega \cdot cm$ is much higher than that reported for the 8 nm Ta layer in Ref. [10], the value of the spin Hall conductivity $\sigma_{SH}^{Ta}$ of the 1 nm Ta layer will be about 3 times smaller than the value in (S9).

Thus we only need to consider Case 1. To estimate the maximum spin current from the Ta layer that reaches the Pt/Co interface, we consider the $t_{Pt} = 1.2$ nm sample (smallest Pt thickness) and assume that all the spin current from the Ta layer flows through the Pt layer without attenuation. The ratio of the spin currents generated by the Ta and Pt layer is, approximately,

$$\frac{j_s^{Ta}}{j_s^{Pt}} \sim \frac{\sigma_{SH}^{Ta}}{\sigma_{SH}^{Pt}} \times \frac{t_{Ta}}{t_{Pt}} \approx -0.12 . \quad\quad\quad (S10)$$

This introduces an error of about $0.06 \times 10^5 / \Omega m$ to the first data point in Fig. 1(d) shown in the main text. In reality, the error caused by the Ta seeding layer will be smaller due to the spin attenuation in the Pt layer.

    The above estimated (maximum) error caused by the 1 nm Ta seeding layer was included in estimating the uncertainty for the data points shown in Fig. 1(c, d) and Fig. 2(b) in the main text. The fitted values reported in the main text were obtained by a weighted fitting technique that takes into account the uncertainty of the data points.

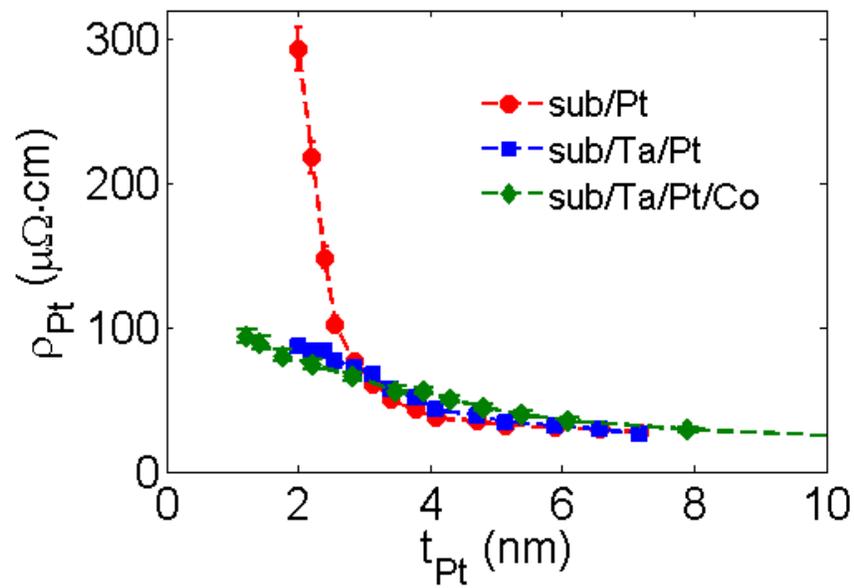

**Figure S3:** Averaged resistivity of Pt in Pt thin films grown directly on Si/SiO$_2$ substrate (red), on 1 nm Ta seeding layer (blue) and in our samples in the main text (green) versus Pt thickness.

**REFERENCES**

[1]     P. C. Van Son, H. Van Kempen, and P. Wyder, Phys. Rev. Lett. **58**, 2271 (1987).

[2]     C. T. Boone, J. M. Shaw, H. T. Nembach, and T. J. Silva, J. Appl. Phys. **117**, 223910 (2015).

[3]     L. Liu, R. A. Buhrman, and D. C. Ralph, arXiv:**1111**.3702 (2011).

[4]     K. Kondou, H. Sukegawa, S. Mitani, K. Tsukagoshi, and S. Kasai, Appl. Phys. Express **5**, 073002 (2012).

[5]     A. Ganguly, K. Kondou, H. Sukegawa, S. Mitani, S. Kasai, Y. Niimi, Y. Otani, and A. Barman, Appl. Phys. Lett. **104**, 072405 (2014).

[6]     W. Zhang, V. Vlaminck, J. E. Pearson, R. Divan, S. D. Bader, and A. Hoffmann, Appl. Phys. Lett. **103**, 242414 (2013).

[7]     C.-F. Pai, Y. Ou, L. H. Vilela-leão, D. C. Ralph, and R. A. Buhrman, Phys. Rev. B **92**, 064426 (2015).

[8]     T. Nan, S. Emori, C. T. Boone, X. Wang, T. M. Oxholm, J. G. Jones, B. M. Howe, G. J. Brown, and N. X. Sun, Phys. Rev. B **91**, 214416 (2015).

[9]     J. M. Shaw, H. T. Nembach, T. J. Silva, S. E. Russek, R. Geiss, C. Jones, N. Clark, T. Leo, and D. J. Smith, Phys. Rev. B **80**, 184419 (2009).

[10]    L. Liu, C.-F. Pai, Y. Li, H. W. Tseng, D. C. Ralph, and R. A. Buhrman, Science **336**, 555 (2012).